\documentclass[aps,twocolumn,showpacs,preprintnumbers,amsmath,amssymb]{revtex4-1}
\usepackage[english]{babel}
\usepackage{color}
\usepackage{float}
\usepackage{amsfonts}
\usepackage{graphicx}
\usepackage{graphics}
\usepackage{subfigure}

\begin{document}

%
%----------------------------head----------------------------------
%
%\title{Exchange Bias in Ferromagnet/Antiferromagnet Bilayered Nanostructures Within an Anisotropic Heisenberg Model}
%\title{Study of the magnetization reversal process in FM/AFM bilayer structures by anisotropic Heisenberg Model}
\title{Spin Dynamics Simulation of the magnetization reversal process in FM/AFM bilayer structures by Anisotropic Heisenberg Model}

\author{E. B. Santos,  R. A. Dias, S. A. Leonel, P. Z. Coura,\\
\small Departamento de F\'{\i}sica, ICE, UFJF, 36036-330 Juiz de Fora, MG, Brazil\\}
\author{J.C.S. Rocha\\
\small Departamento de F\'{\i}sica, ICEX, UFMG, 30123-970, Belo Horizonte, MG, Brazil}
\begin{abstract}
We have studied the magnetization reversal process in FM/AFM bilayer structures through of  spin dynamics simulation. It has been observed that the magnetization behavior is different at each branch of the hysteresis loop as well as the exchange-bias behavior. On the descending branch a sudden change of the magnetization is observed while on the  ascending branch is observed a bland change of the magnetization. The occurrence of the asymmetry in the hysteresis loop and the variation in the exchange-bias is due to anisotropy which is introduced only in the coupling between ferromagnetic (FM) and antiferromagnetic (AFM) layers. 
\end{abstract}
%
%\begin{keyword}
%exchange bias, magnetic nano-disk, Monte Carlo, Heisenberg, dipolar interaction
%\end{keyword}
%
\maketitle
%
%----------------------------Section----------------------------------
%
\section{Introduction}

In the last decades much research into thin magnetic films has been driven by several interesting properties associated with interfaces involving magnetic materials. The key to the understand these magnetic structures is the knowledge about the basic energies involved. There are three basic energy concepts: exchange, dipolar and anisotropy energies. The first controls the magnetic ordering, the second controls the form of the magnetic structures and the latter controls the preferred orientation. Both are phenomenological descriptions of fundamental correlations and energies associated with the electronic and crystalline structure of the material. We must also take into account the surface effects  on the magnetic properties because they become increasingly important when the particle size decreases. Systems with reduced dimensionality show properties different from the bulk materials due to finite-size effects. In particular, magnetic ordering temperature and magnetocrystalline anisotropy were shown to be size dependent in many materials.
One of the most important effects studied is the Exchange-Bias(EB) effect, discovered in 1956 \cite{primeiro,melkojohn1962}. The EB effect occurs due to exchange coupling between ferromagnetic(FM) and antiferromagnetic(AFM). Its signature is a shifted hysteresis loop dislocated from zero field. The shift is attributed to the frozen in global unidirectional anisotropy of the system. Several interesting properties are obtained from hysteresis loop as the remanent magnetization which is a residual magnetization in the material when the aligning field is reduced to zero, the coercive field required for canceling the residual magnetization and the saturation field which is the field required for forcing all magnetic moments of the sample to point in the direction of the external field. These features make the exchange bias effect an efficient and reliable way for pinning ferromagnetic layers in spin-valves. Consequently it is largely exploited in technological applications, such as magnetic sensors and data storage devices. At the same time, several aspects of this effect are still unclear from a microscopic point of view, mostly due to different exchange coupling between ferromagnetic(FM) and antiferromagnetic(AFM) layers and it  is difficult to obtain a unique theory \cite{review1, review2, review3,review4,review5,experimental1,experimental2}.

In this work we report a Spin Dynamics Simulation to study the magnetic behavior of a FM layer put over the (100) face of an uncompensated\cite{review3,review4,review5} AFM substrate both with a BCC structure. The scope of this article is to discuss the role of the exchange interaction across the interface when we insert an anisotropy in the classical XYZ Heisenberg model to calculate the interactions  between the magnetic moments  in the FM/AFM interface. In the next session we show our theoretical model.

%----------------------------Section----------------------------------
%

\section{Model}

\begin{figure}[h!]
\centerline{\resizebox*{9.0
cm}{!}{\includegraphics{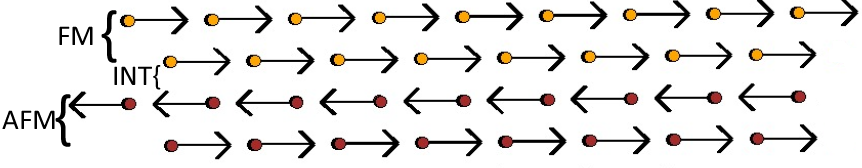}}}
\caption{Schematic representation of the system. Figure shows FM layers put over an AFM substrate and the region INT between them. The plane of the figure is the $zx$-plane ( the $x$-axis is pointing to right, the $y$-axis is pointing into the paper and $z$-axis is pointing upward) and the total number of FM and AFM particles is given respectively by $n_{x}\times n_{y}\times n^{FM}_{z}$ and $n_{x}\times n_{y}\times n^{AFM}_{z}$. In this work the number of particles in the plane, $n_{xy} = n_x \times n_y = 15 \times 15 = 225 $, is the same to FM and AFM.}
\label{system1}
\end{figure}

Our system was divided into three regions: FM region, AFM region and the INT region (FM/AFM interface). The schematic view is represented in the figure \ref{system1}.
%%%%%%%%%%%%%%%%%%%%%%%%%%%%%%%
%%%%%%%%%%%%%%%%%%%%%%%%%%%%%%

Theoretically we can consider a pseudo-spin Hamiltonian model to describe the system. In this work we use the following Hamiltonian \cite{kiwi,jhseok,shanho,jmazo,julio}

%%%%%%%%%%%%%%%%%%%%%%%%%%%%%%
\begin{eqnarray}
& H &  =  -J_{FM} \sum_{<i,j>~ \varepsilon~FM} ( S_i^x S_j^x + S_i^y S_j^y + S_i^z S_j^z) - \nonumber  \\ 
& - & g_{FM} \mu_{B}\sum_{<i>~ \varepsilon~FM} \vec{S}_i \cdot \vec{H}_{ex} - \nonumber \\
 & - &  J_{AFM} \sum_{<i,j>~ \varepsilon~AFM} ( S_i^x S_j^x + S_i^y S_j^y + S_i^z S_j^z) - \nonumber \\
 & -  & g_{AFM}\mu_{B}\sum_{<i>~ \varepsilon~AFM} \vec{S}_i \cdot \vec{H}_{ex} +  \nonumber  \\
 & + &  K \sum_{<i>~ \varepsilon~AFM} [1-(\vec S_i^{x})^2 ] \label{hamiltonian}  - \nonumber \\
 & - & J_{INT}\sum_{<i,j>~ \varepsilon~INT} (\lambda_x S_i^x S_j^x + \lambda_y S_i^y S_j^y + \lambda_z S_i^z S_j^z) 
 \label{hamiltoniana}
 \end{eqnarray}
%%%%%%%%%%%%%%%%%%%%%%%%%%%%%%%
 
where $S_i^{(x,y,z)}$ denotes the spin components at lattice site $i$ with the normalization $|\vec S_{i}| = 1$, $S_j^{(x,y,z)}$ denotes the neighbor spin, $J_{FM}$ is the exchange coupling constant for FM, $g_{FM}=g_{AFM}=g$ is the giromagnetic ratio for FM and AFM, $J_{AFM}=-0.1J_{FM}$ is the exchange coupling constant for AFM, $K=10J_{FM}$ is the uniaxial anisotropy constant for AFM in the x-axis, $\vec H_{ex}$ is the external magnetic field with $-1.6 J_{FM}/g\mu_b \leq H_{ex} \leq1.6 J_{FM}/g\mu_b$ and $\mu_{B}$ is the Bohr`s magneton. The constants  $J_{INT}$, $\lambda_x,\lambda_y$, and $\lambda_z$ describes the strength of the exchange coupling anisotropy in the FM/AFM interface. Without loss of generality we can set $J_{FM}=1$, thus the energy of system will be given as a function of $J_{FM}$. For typical value $J_{FM}=1$meV, the $H_{eb}$ value is about of $0.1$T or $1$kOe.
\\
We restrict our attention to the case where $-1 \leq J_{INT} \leq 0$, $\lambda_x = 1$, $ 0 \leq \lambda_y \leq 2$, and $0 \leq \lambda_z \leq 2$. Therefore, the interaction in the interface is of the kind antiferromagnetic with a anisotropic behavior. The choose those parameters will determine an easy-axis or easy-plane behavior\cite{jer,saleonel,hanslandau,rtsfreire}. Looking the equation \ref{hamiltonian}, we can note that when $\lambda_x > \lambda_y = \lambda_z$, the system has a easy-axis behavior and the energy is minimized in x-direction. If $\lambda_x < \lambda_y= \lambda_z$, the behavior is easy-plane and the energy is minimized when the  magnetization is in the yz-plane. When $\lambda_x = \lambda_y = \lambda_z=1$ we have an isotropic Heisenberg model. 
\\
Spin dynamics can be described phenomenologically through the well-known Landau-Lifshitz equation \cite{llequation}:
%%%%%%%%%%%%%%%%%%%%%%%%%%%%%%%%%%
\begin{equation}
\frac{ d\vec S}{dt}=\vec{S}_i \times [\vec{H}_{eff}(t) - \alpha \vec{S}_i \times \vec{H}_{eff}(t) ],
\label{eqdemov}
\end{equation}
%%%%%%%%%%%%%%%%%%%%%%%%%%%%%%%%%%
where
%%%%%%%%%%%%%%%%%%%%%%%%%%%%%%%%%%
\begin{equation}
\vec{H}_{eff}(t)= - \frac{\partial H}{\partial \vec{S}_i}.
\end{equation}
%%%%%%%%%%%%%%%%%%%%%%%%%%%%%%%%%%
 and  $\alpha$ is the damping parameter. In this work we use $\alpha = 0.05$ and  the Adams-Moulton predictor-corrector method \cite{Adams,Adams1} to solve this equation.

%\pagebreak
%
\section{Simulation results}
\subsection{Isotropic Heisenberg Model}
\subsubsection{Hysteresis}

\begin{figure}[!h]
\subfigure[$n^{FM}_{z} = 2$]{
\includegraphics[width=7.2cm]{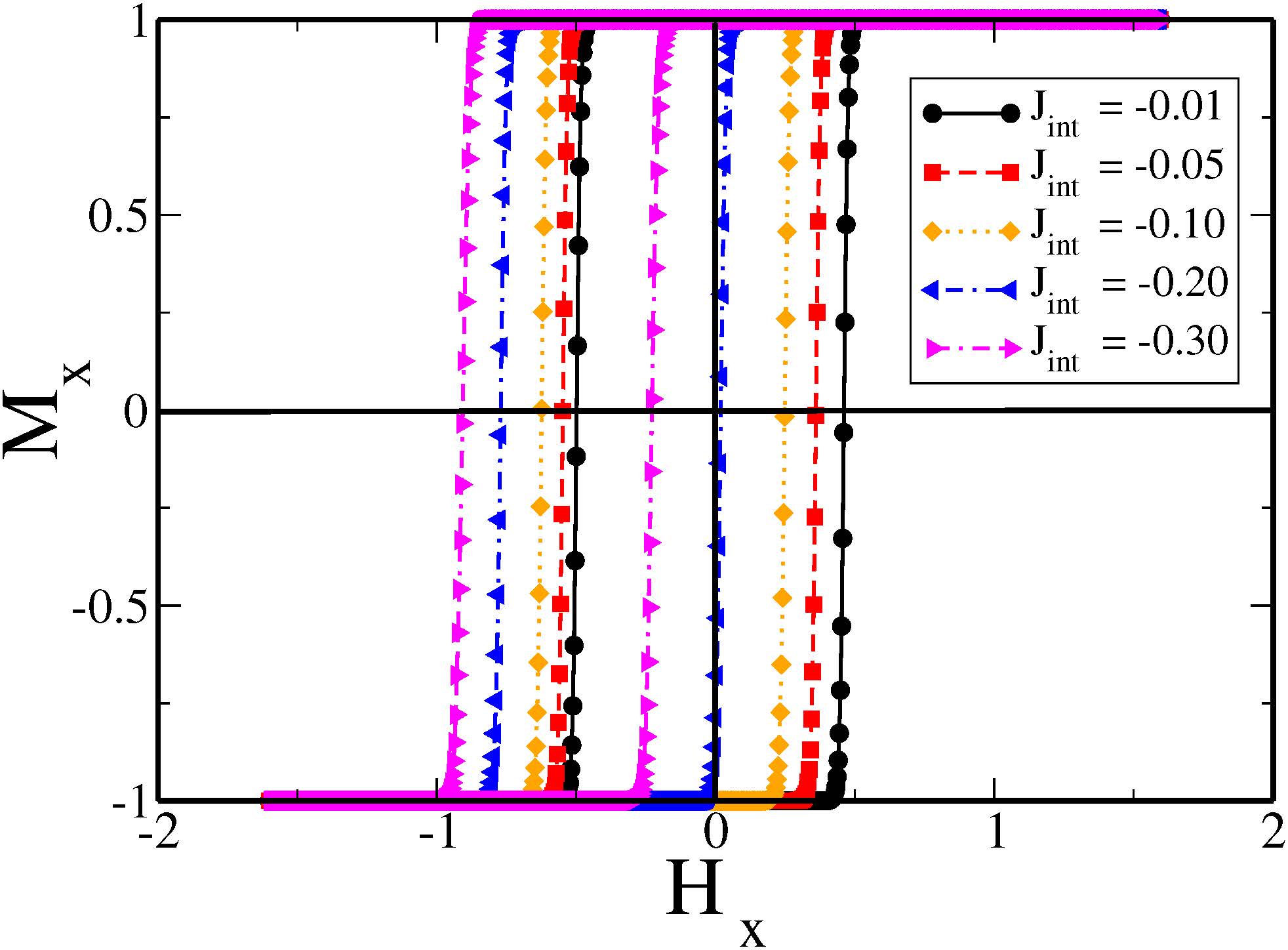}
\label{his_iso1800}
}
\quad 
\subfigure[$n^{FM}_{z} = 6$]{
\includegraphics[width=7.2cm]{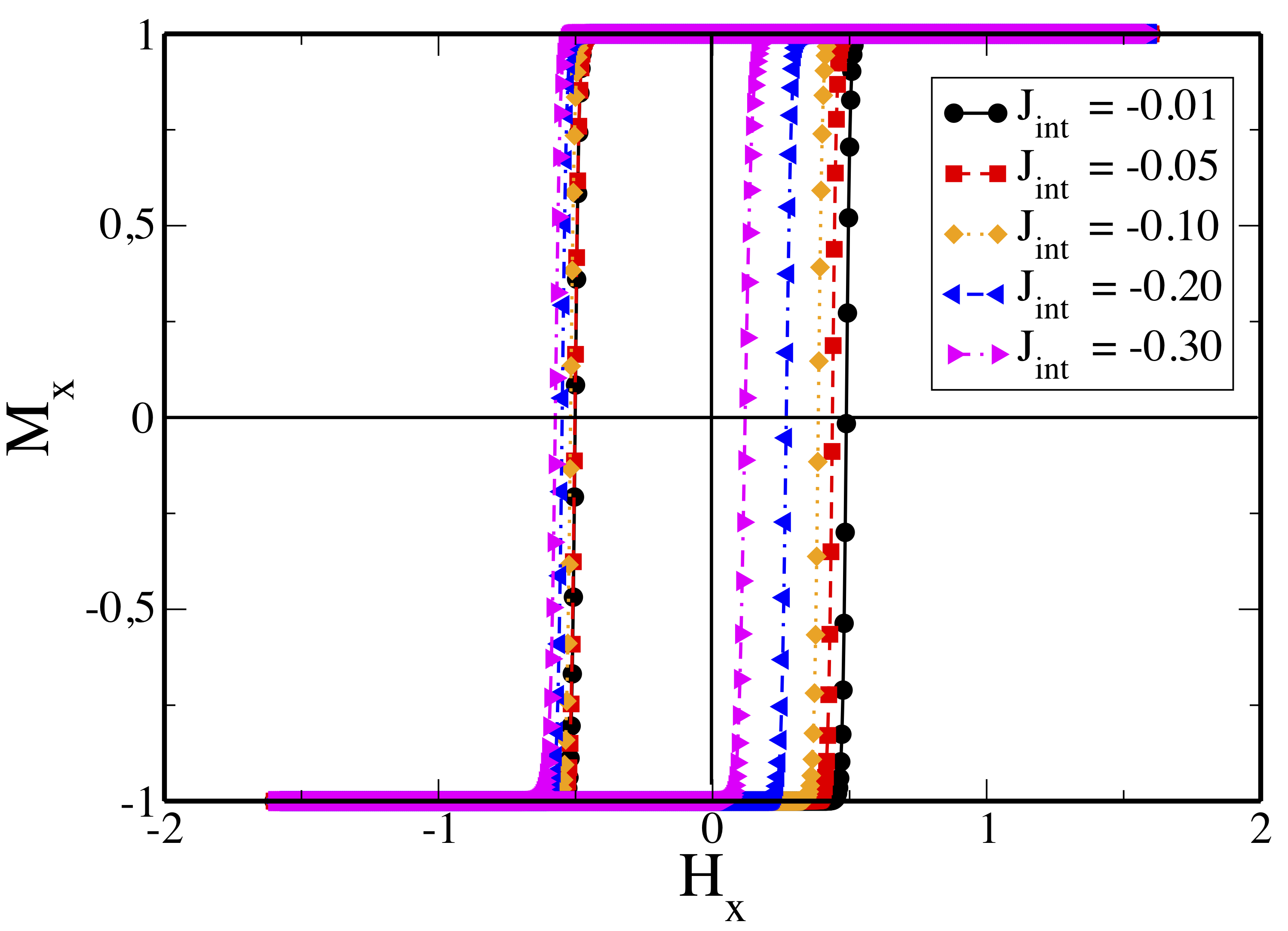}
\label{his_iso2700}
}
\caption{Hysteresis loop of the magnetization $M_{x}$ as function of $H_{x}$ to 2 and 6 FM layers. The results to $J_{INT} < -0.30$ is not shown here.}
\label{his_iso}
\end{figure}

 In this section we show our results for the isotropic Heisenberg model. We present in figure \ref{his_iso} the hysteresis loop for the $x$-component of magnetization ($M_{x}=\frac {\sum\limits_{<i> \in FM} S^{x}_i}{n_{x}\times n_{y} \times n^{FM}_{z}} $). In this figure the total number of FM and AFM layers are, respectively, $n^{FM}_{z} = 2$ and $n^{AFM}_{z}= 6$ in figure \ref{his_iso1800}, and $n^{FM}_{z} = 6$ and $n^{AFM}_{z}= 6$ in figure \ref{his_iso2700}. The numerical calculation is done as follows: An initial configuration is chosen as shown in the figure \ref{system1}. Then an external magnetic field is applied in the $x$-direction ($H_x$). This field is turned on at time $t = 0$ and changed in steps of size $\Delta H = 0.02$. For each value of the magnetic field $5\times 10^{2}$ time steps are performed in order to equilibrate the system. When the FM layers reached the saturation the magnetic field is increased or decreased until these layers saturates in the inverse direction.

%%%%%%%%%%%%%%%%%%%%%%%%%%%%%%%%%%
%%%%%%%%%%%%%%%%%%%%%%%%%%%%%%%%%%

It is observed that increasing the number of  FM layers, the EB effect decreases. These behaviors are in accordance with the results found in the literature \cite{kiwi}. Using the results from the figure \ref{his_iso} we can measure the exchange-bias field $H_{eb}$ and the coercive field $H_{c}$ which are shown in next section.
\\
\subsubsection{Exchange-bias field $H_{eb}$ and coercive field $H_{c}$}

\begin{figure}[!h]
\subfigure[$n^{FM}_{z} = 2$]{
\includegraphics[width=7.2cm]{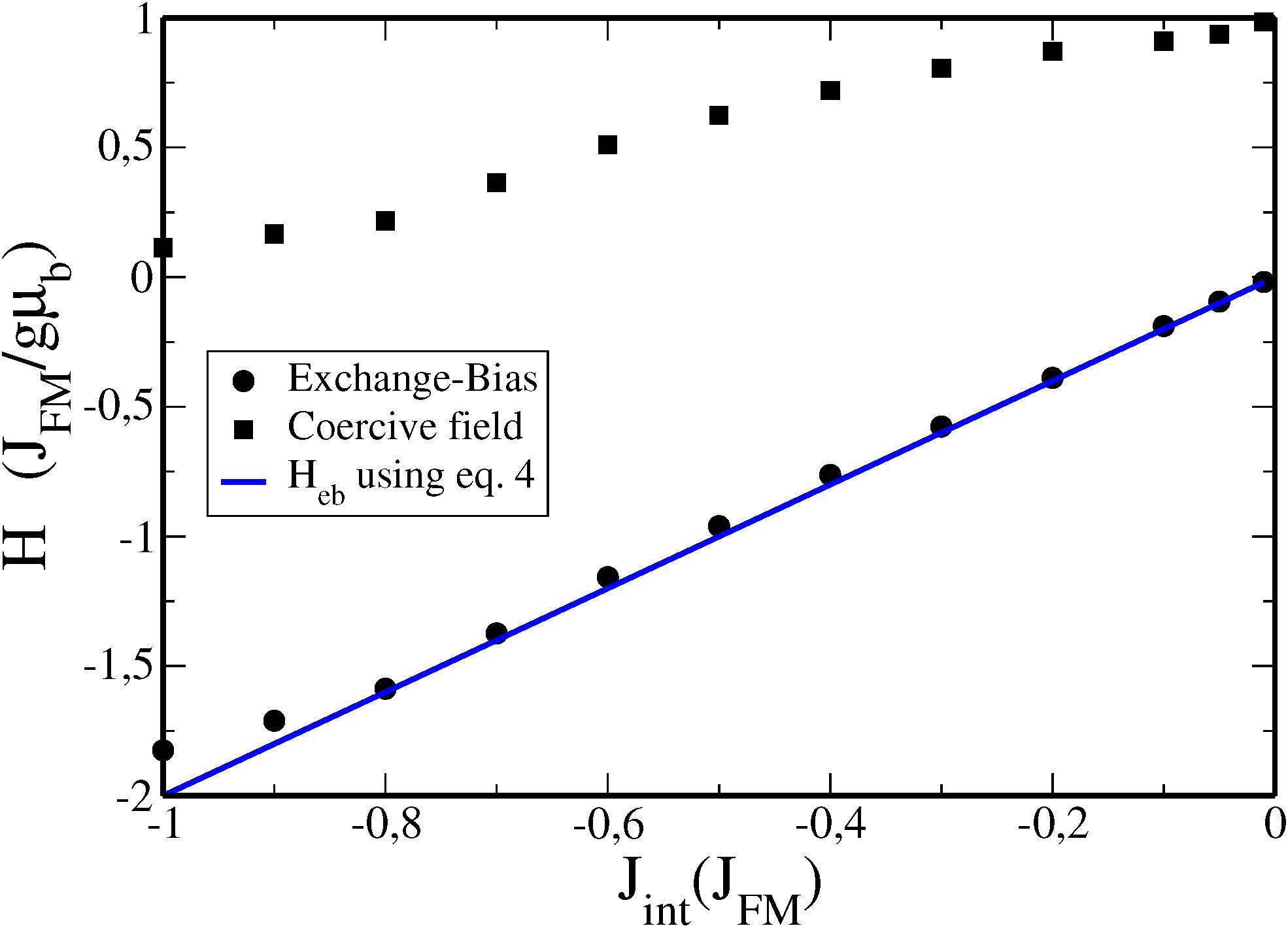}
\label{hebxjac1800}
}
\quad %espaco separador
\quad %espaco separador
\subfigure[$n^{FM}_{z} = 6$]{
\includegraphics[width=7.2cm]{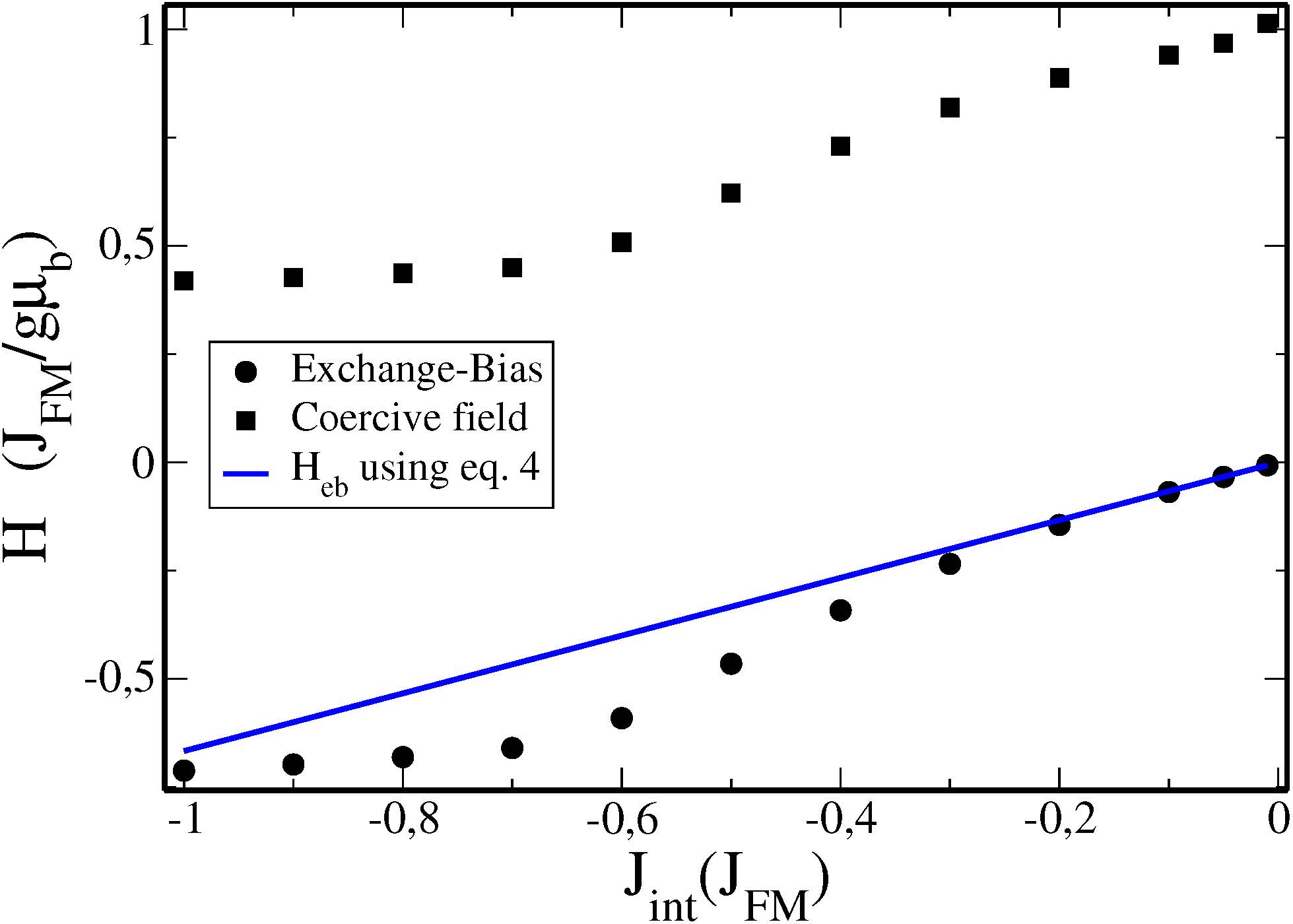}
\label{hebxjac2700}
}
\caption{$H_{eb}$ and $H_{c}$ as function of $J_{INT}$ for isotropic system to 2 and 6 FM layers.}
\end{figure}

According to the model of Meiklejohn-Bean\cite{orlando} the exchange-bias field is given by
%%%%%%%%%%%%%%%%%%%%%%%%%%%%%%%%%%
\begin{equation}
H_{eb}=\frac{4 J_{INT}} {n^{FM}_z}.
\label{teoricozuntim}
\end{equation}
%%%%%%%%%%%%%%%%%%%%%%%%%%%%%%%%%%
Figures \ref{hebxjac1800} ($n^{FM}_{z} = 2$) and figure \ref{hebxjac2700} ($n^{FM}_{z} = 6$) show our results to  $H_{eb}$ and $H_{c}$ as function of $J_{INT}$. As can be observed in the figure \ref{hebxjac1800}, the behavior of the $H_{eb}$ as a function of the $J_{INT}$ is linear for values up to approximately $J_{INT} = -0.8$. Note that this results are consistent with the theoretical result (continuous line).
\\
%%%%%%%%%%%%%%%%%%%%%%%%%%%%%%%%%%
%%%%%%%%%%%%%%%%%%%%%%%%%%%%%%%%%%
 However, in the figure \ref{hebxjac2700}, the expected linear behavior is obtained just for small values of the $| J_{INT}| $.
Among various conditions it is assumed in Meiklejohn-Bean's approach that the magnetization of the FM layer rotates rigidly.
This condition is satisfied just for thin films. For thick films this condition is not fulfill since each FM layer present a different magnetization direction during the rotation.
We can understand it looking at the EB effect mechanism. This effect is due to an effective field caused by the FM/AFM interface.
This effective field depends mainly on the direction of the spins and the exchange interactions on the interface.
It lose strength on uppers layers (far away from the interface) due the low-range caracter of the exchange interaction.
It means that each FM layer ``sees'' a different magnetic field.
Therefore the condition of rigid rotation can not be fulfill since the effective local field is not the same for all FM layers.
This loosely rotation also explain the difference of the dependence of the coercive field behavior with $J_{INT}$ on the applied field direction. 
For negative external field $H_c$ is almost the same for all $J_{INT}$ values. For the positive external field one can observe a dependence on $H_c$ with $J_{INT}$.
Once the effective field points on $+x$-direction it will contribute (compete) with positive (negative) external field.
For negative external field the magnetization of the uppers layers gyrates first than the lowers one.
For thick films uppers layers domain the total magnetization, so the effect due the interface is negligible.
On the other side, due the contribution of the fields, the effect due the interface can be easily propagate through the FM layer for positive external field.
Of course not as easily as on thin films. However in the limit of bulk materials one can overlook interface effects and EB no longer happens.
\\
%%%%%%%%%%%%%%%%%%%%%%%%%%%%%%%%%%
\\
\subsection{Anisotropic Heisenberg Model}

In this section we show our results for the anisotropic Heisenberg model. We set $\lambda_x = 1$ and varied the parameters $J_{INT}$, $\lambda_y$ and $\lambda_z$. All results  was obtained to a system of dimensions $n_{xy}= 225 $ and $n^{AFM}_{z} = n^{FM}_{z} = 6$.

\subsubsection{Exchange-bias field $H_{eb}$ and coercive field $H_{c}$}

\begin{figure}[!h]
\subfigure[$\lambda_x=1$, $\lambda_z=0$]{
\includegraphics[width=7.2cm]{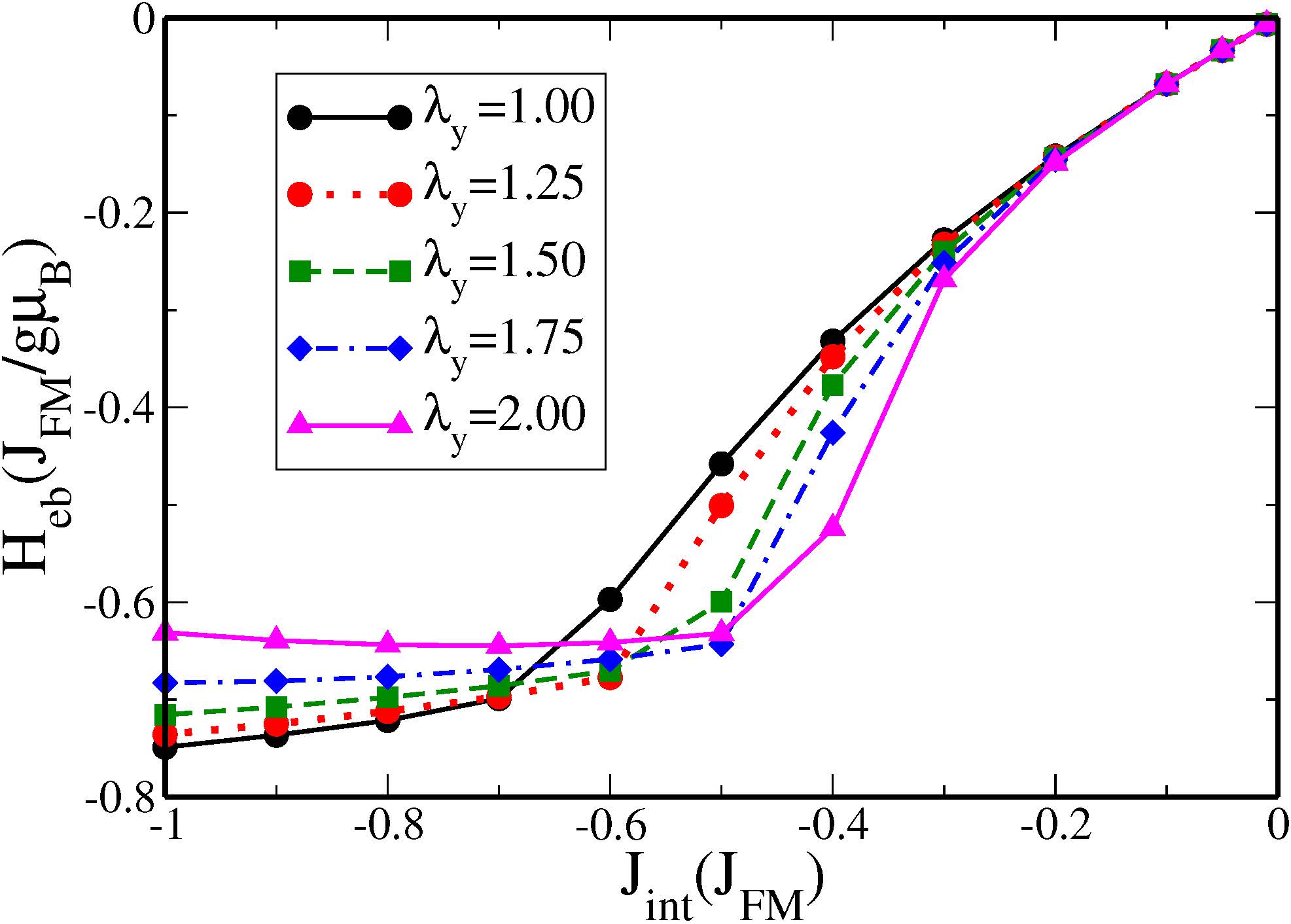}
\label{heb_x_jint_lambz0}
}
\quad %espaco separador
\quad %espaco separador
\subfigure[$\lambda_x=1$, $\lambda_y=0$]{
\includegraphics[width=7.2cm]{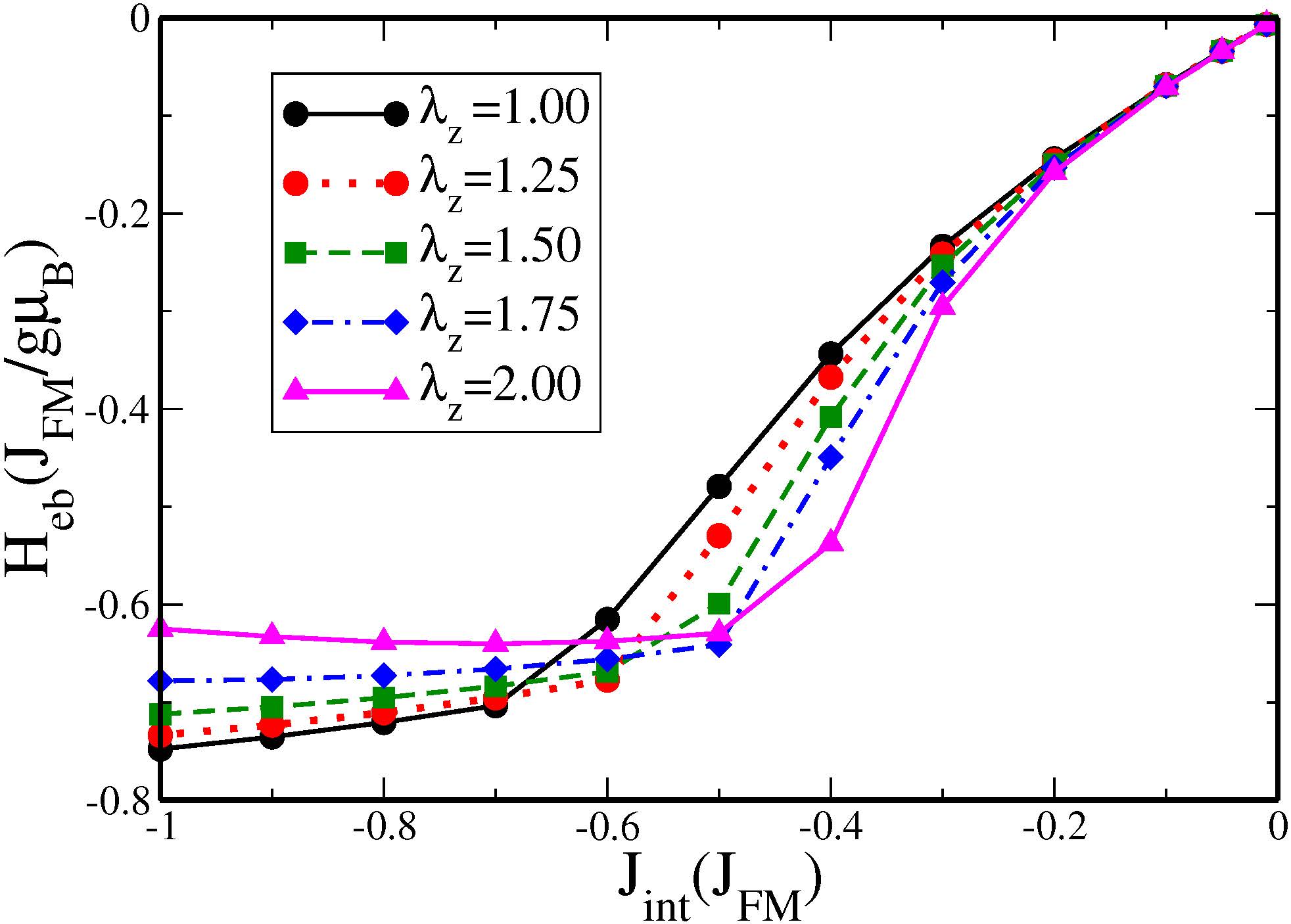}
\label{heb_x_jint_lamby0}
}
\caption{Exchange-Bias field as function of $J_{INT}$ for several values of $\lambda_y$ and $\lambda_z$ .}
\label{heb_x_jint}
\end{figure}

\begin{figure}[!h]
\subfigure[$\lambda_x=1$, $\lambda_z=0$]{
\includegraphics[width=7.2cm]{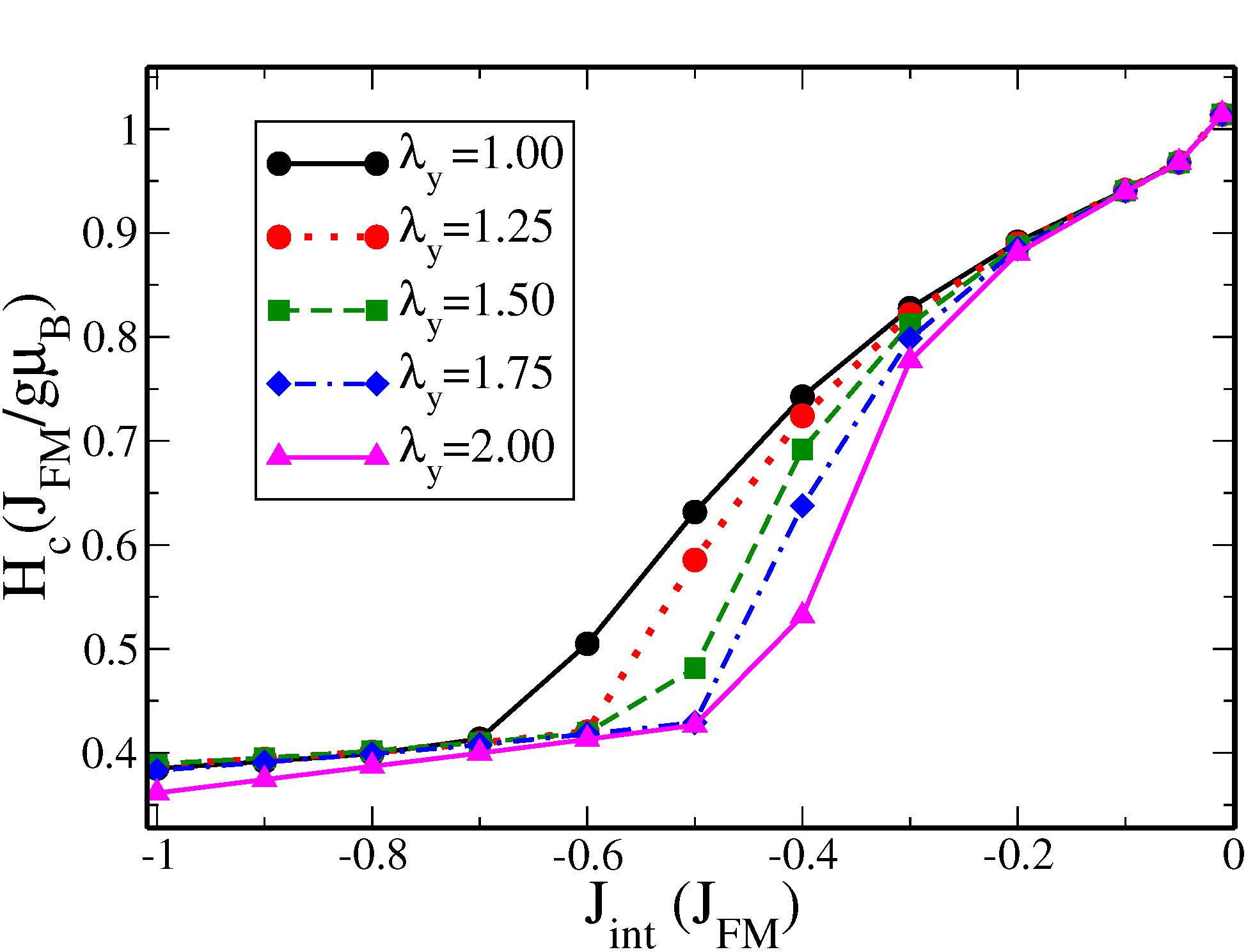}
\label{hc_x_jint_lambz0}
}
\quad %espaco separador
\quad %espaco separador
\subfigure[$\lambda_x=1$, $\lambda_y=0$]{
\includegraphics[width=7.2cm]{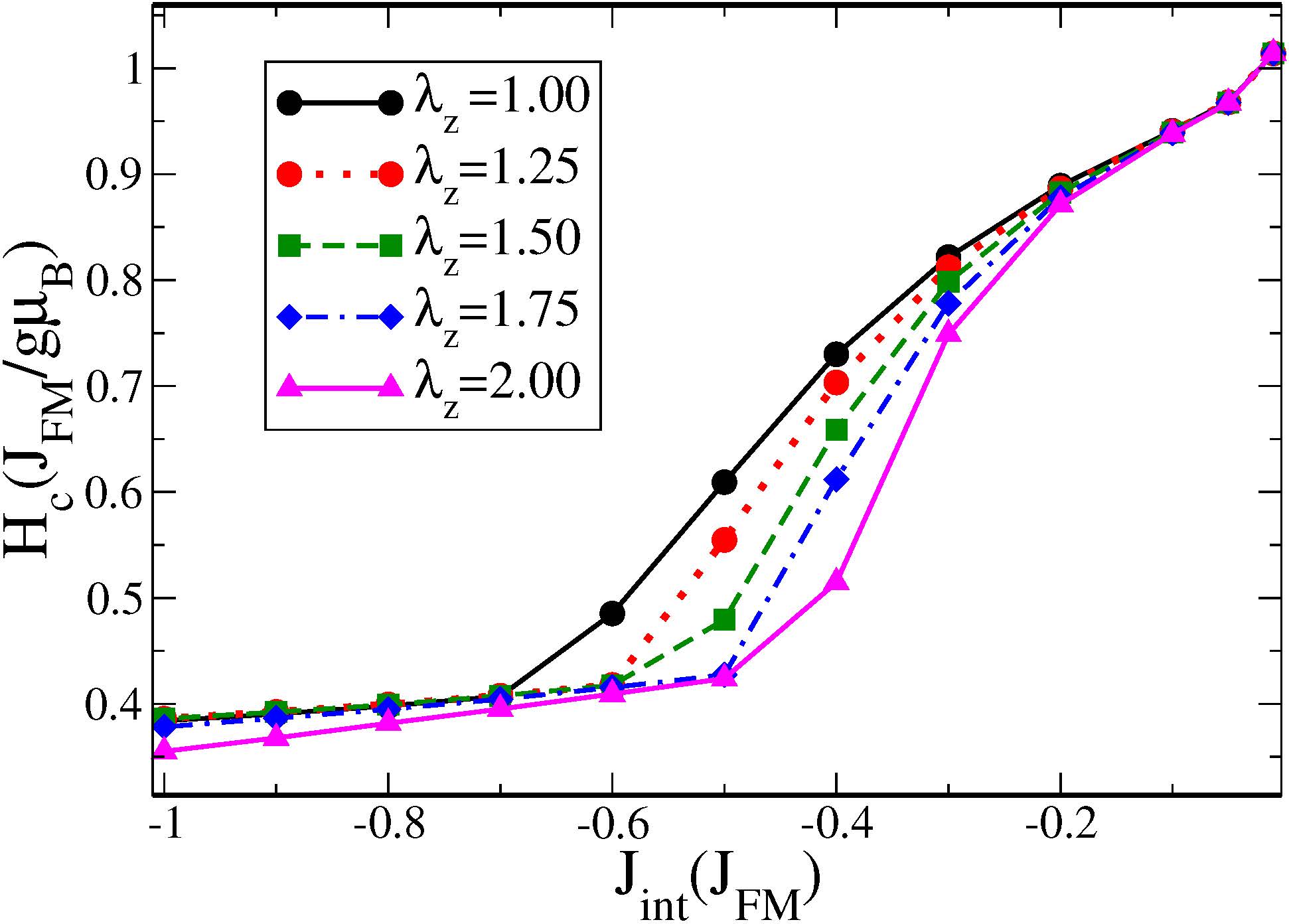}
\label{hc_x_jint_lamby0}
}
\caption{Ccoercive field as function of $J_{int}$, for several values of  $\lambda_y$ and $\lambda_z$ .}
\label{hc_x_jint}
\end{figure}

We present in figure \ref{heb_x_jint} our results of $H_{eb}$ as function of $J_{INT}$ to several values of $\lambda_y$ and $\lambda_z$.
In figure \ref{heb_x_jint_lambz0} (figure \ref{heb_x_jint_lamby0}) we set $\lambda_x=1$ and $\lambda_z=0$ ($\lambda_x=1$ and $\lambda_y=0$) and varied $\lambda_y$ ($\lambda_z$) in order to study the anisotropy parallel (perpendicular) to the interface.
For $\lambda_x = \lambda_y$ ($\lambda_x = \lambda_z$) we have an easy-plane anisotropy which favors the spin alignment in the xy-plane (xz-plane).
Increasing $\lambda_y$ ($\lambda_z$) this anisotropy tends to behavior like an easy-axis one in the $y$-direction ($z$-direction).
Due the high uniaxial anisotropy on the substrate the AFM moments are almost stuck on $x$-direction. 
So this direction dominates the interface exchange. Between this and that the effective field at the interface have a high component on $x$-direction.
During the process of reversing the magnetization was observed a small component of the effective field in other directions.
The anisotropy is important in this process. In order to minimize the energy, if we have an easy-plane anisotropy this component can point in any direction of the plane.
On the other hand, if we have an easy-axis anisotropy this component point in the direction of the easy-axis.
The combination of the external field and the effective field determines the direction that the magnetics moments will
precess around, during the inversion of the magnetization. This is a coupled process, when the spins precess it also change the axis of precession.
We are going to talk about this precession on the next section.
Thereby there is no difference considering $xy$ or $xz$ easy-axis anisotropy, neither $y$ or $z$ as easy-axis anisotropy, as can be seen.
We show in figure \ref{hc_x_jint} our results for the coercive field $H_c$.
One can observe that, for intermediate $J_{INT}$ values, $H_c$ decreases when anisotropy parameter $\lambda_y$ or $\lambda_z$ increases.
We are going to explore this region on the next section.
\\
%%%%%%%%%%%%%%%%%%%%%%%%%%%%%%%%%%
\\
%%%%%%%%%%%%%%%%%%%%%%%%%%%%%%%%%%

%%%%%%%%%%%%%%%%%%%%%%%%%%%%%%%%%%

\begin{figure}[h!]
\centerline{\resizebox*{7.0
cm}{!}{\includegraphics{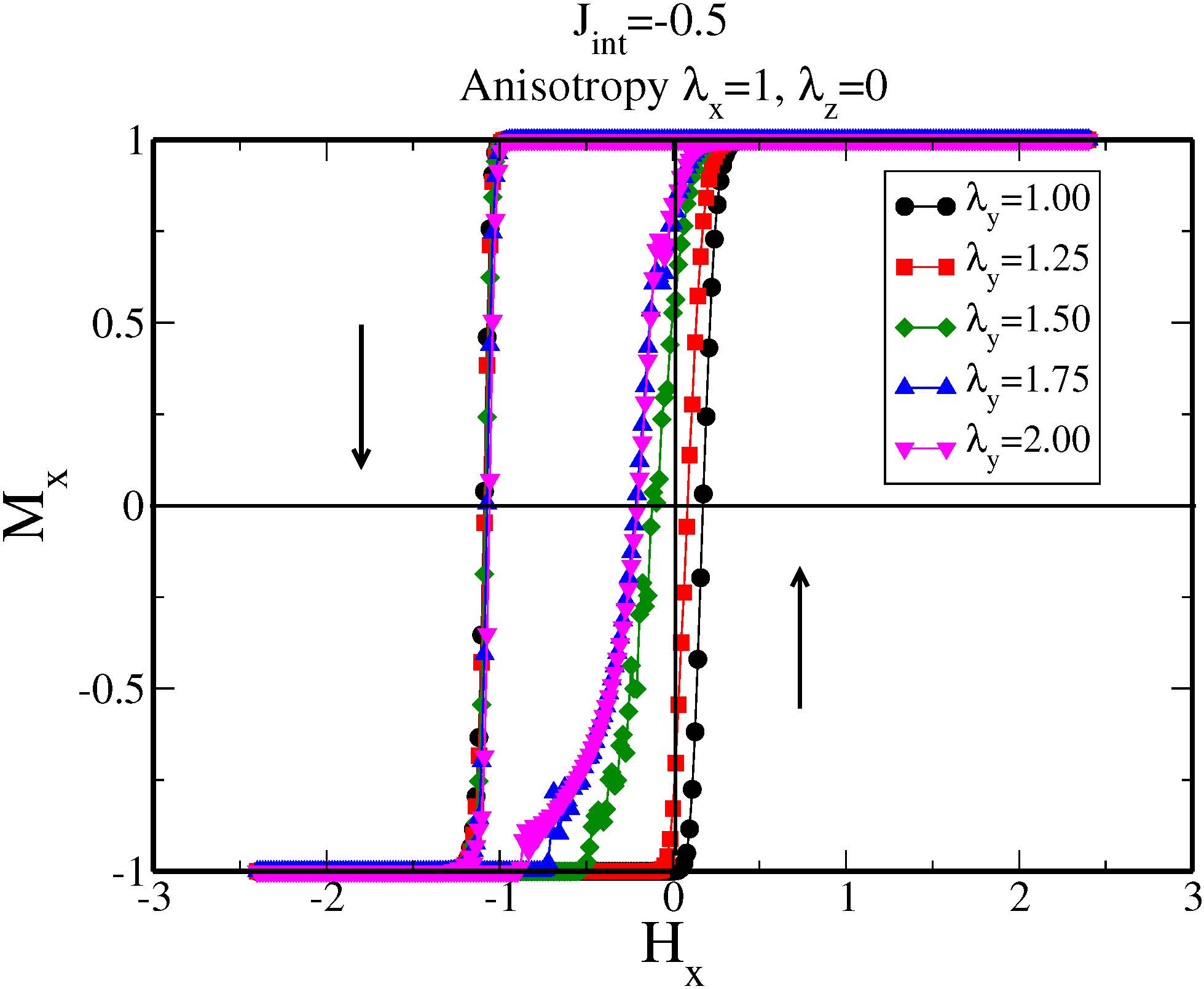}}}
\caption{Hysteresis loop for several values of $\lambda_y$ to $J_{int}=-0.5$.}
\label{histereses_lamby_jint=0.5}
\end{figure}
%%%%%%%%%%%%%%%%%%%%%%%%%%%%%%%%%%

Figure \ref{histereses_lamby_jint=0.5} shows the hysteresis loop to $J_{int}=-0.5$, $\lambda_x = 1$, $\lambda_z = 0$ and $1 \leq \lambda_{y} \leq 2$.  How can be observed the hysteresis loop becomes asymmetrical in the ascending magnetization. The result (not shown) is similar when $\lambda_x = 1$, $\lambda_{y}= 0$ and $1 \leq \lambda_{z} \leq 2$ or $\lambda_y = 1$, $\lambda_x = 0$ and $1 \leq \lambda_{z} \leq 2$  or $\lambda_z = 1$, $\lambda_x = 1$, and $1 \leq \lambda_{y} \leq 2$.

\subsubsection{Asymmetric hysteresis }

\begin{figure}
\centerline{\resizebox*{8.5cm}{!}{\includegraphics{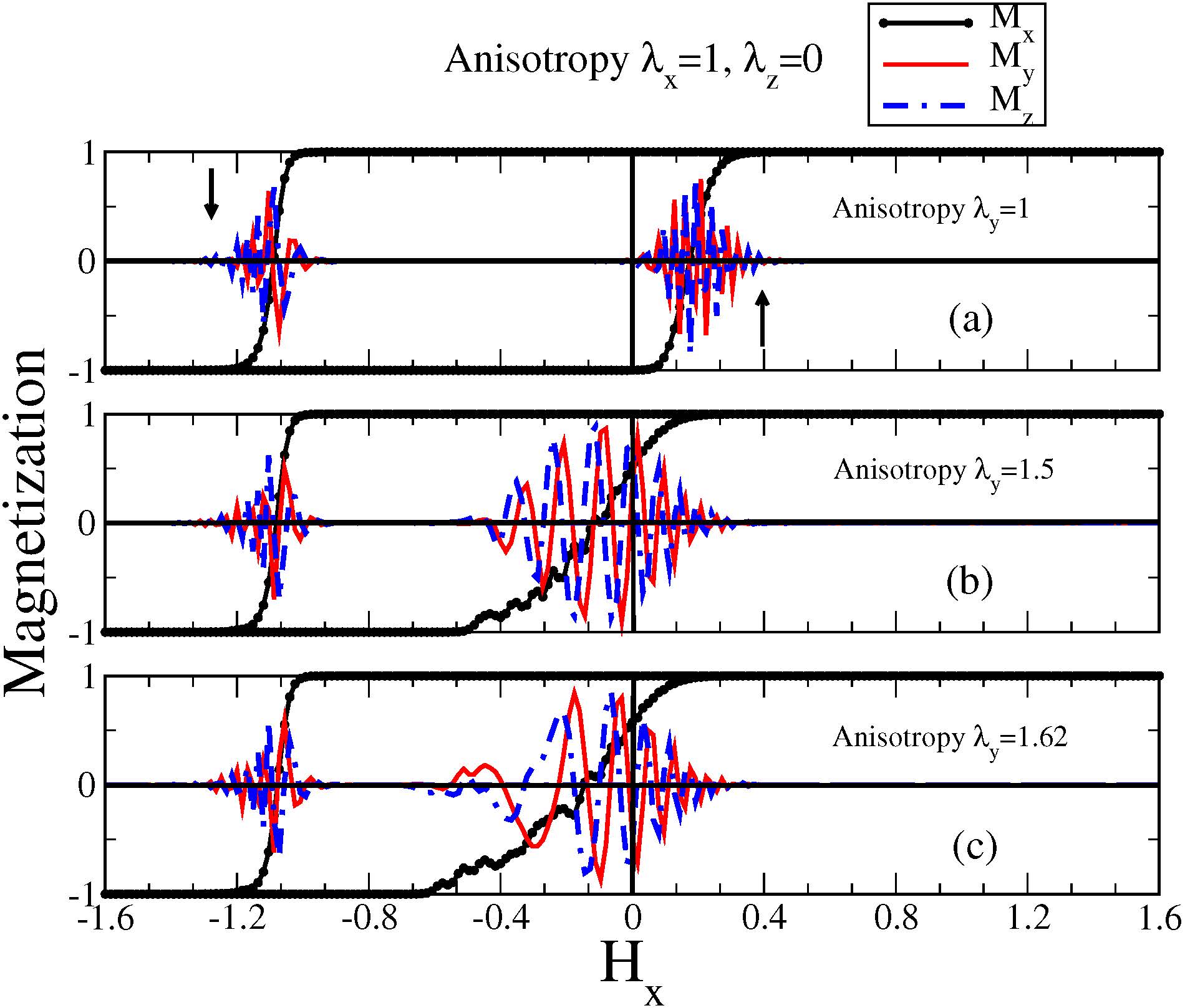}}}
\caption{Magnetization in the directions $x$, $y$ and $z$ to $1 \leq \lambda_{y} \leq 1.62$. It is noticed a change in the coherent magnetization precession  when increasing $\lambda_{y}$. The precession is around the $x$-direction and the precession rate is altered by exchange coupling anisotropy.  }
\label{histereses_new_lambz}
\end{figure}

The hysteresis loop becomes asymmetric when the value  of the anisotropy is increased. Analysing the figure \ref{histereses_new_lambz} we can understand the magnetization behavior during the hysteresis loop. The black curve is the magnetization in the $x$-direction ($M_x$), the red curve is the magnetization in the $y$-direction ($M_y$) and the dashed blue curve is the magnetization in the $z$-direction ($M_z$). Figure \ref{histereses_new_lambz}$(a)$ shows the magnetization for a system without anisotropy in the $xy$-plane and we can observe the coherent magnetization precession $M_y$ e $M_z$ for fields close to coercive field. 

We identified two distinct behavior which cause asymmetric loop. First, to $1 \leq \lambda_{y} \leq 1.62$ (figure \ref{histereses_new_lambz}), the  magnetization filed oscillates around the x-direction drawing a cone whose base radius increase proportionately to value of anisotropy. This change causes the magnetization to spiral in toward the effective field altering the precession rate.
%%%%%%%%%%%%%%%%%%%%%%%%%%%%%%%%%%
%%%%%%%%%%%%%%%%%%%%%%%%%%%%%%%%%%

%%%%%%%%%%%%%%%%%%%%%%%%%%%%%%%%%%
\begin{figure}[h!]
\centerline{\resizebox*{8.5cm}{!}{\includegraphics{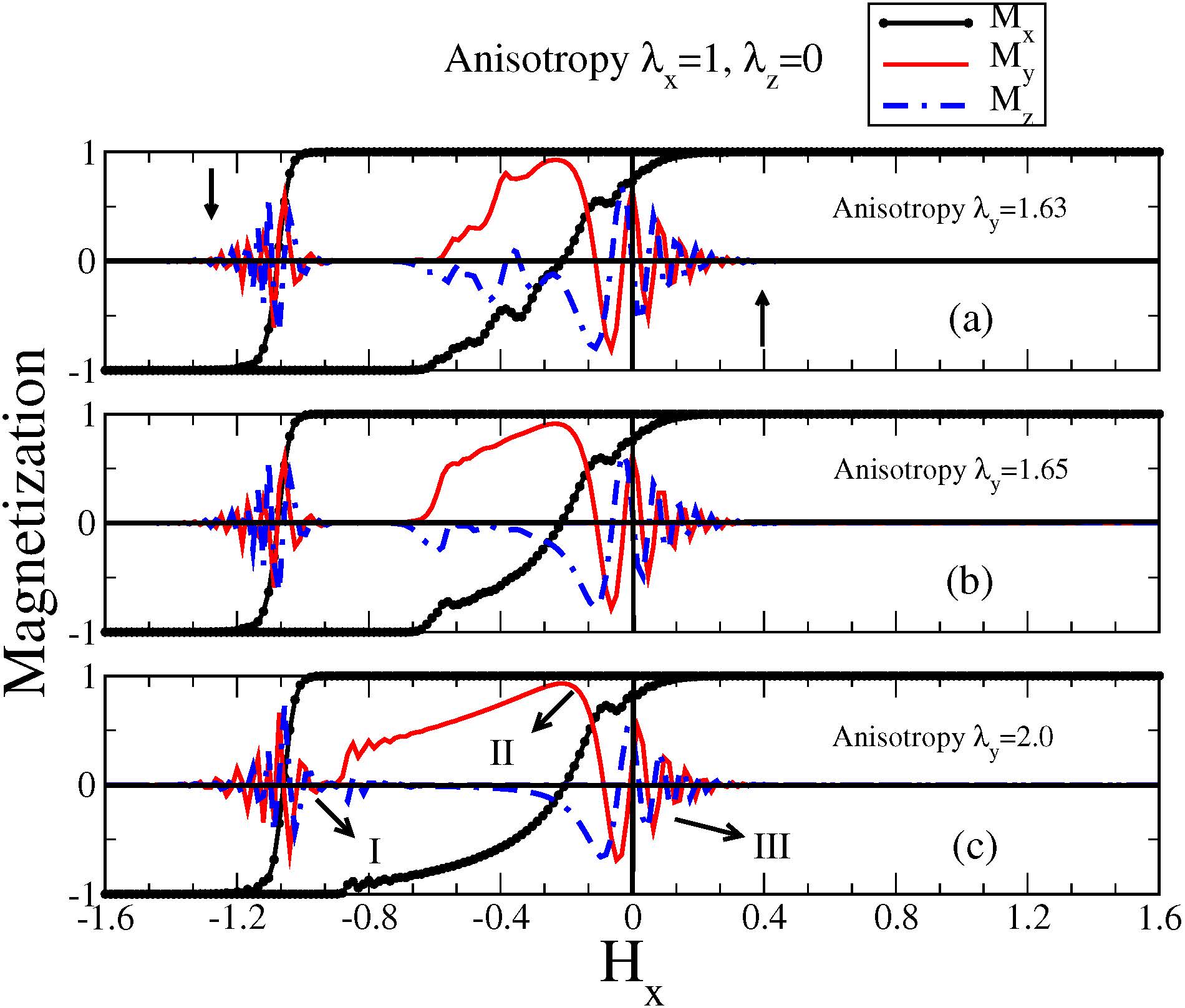}}}
\caption{Magnetization in the directions $x$, $y$ and $z$. The values of the anisotropy are shown in the figure. The precession is around the $z$-direction.}
\label{histereses_new_163_lambz}
\end{figure}
%%%%%%%%%%%%%%%%%%%%%%%%%%%%%%%%%%
\begin{figure}
\includegraphics[width=7.2cm]{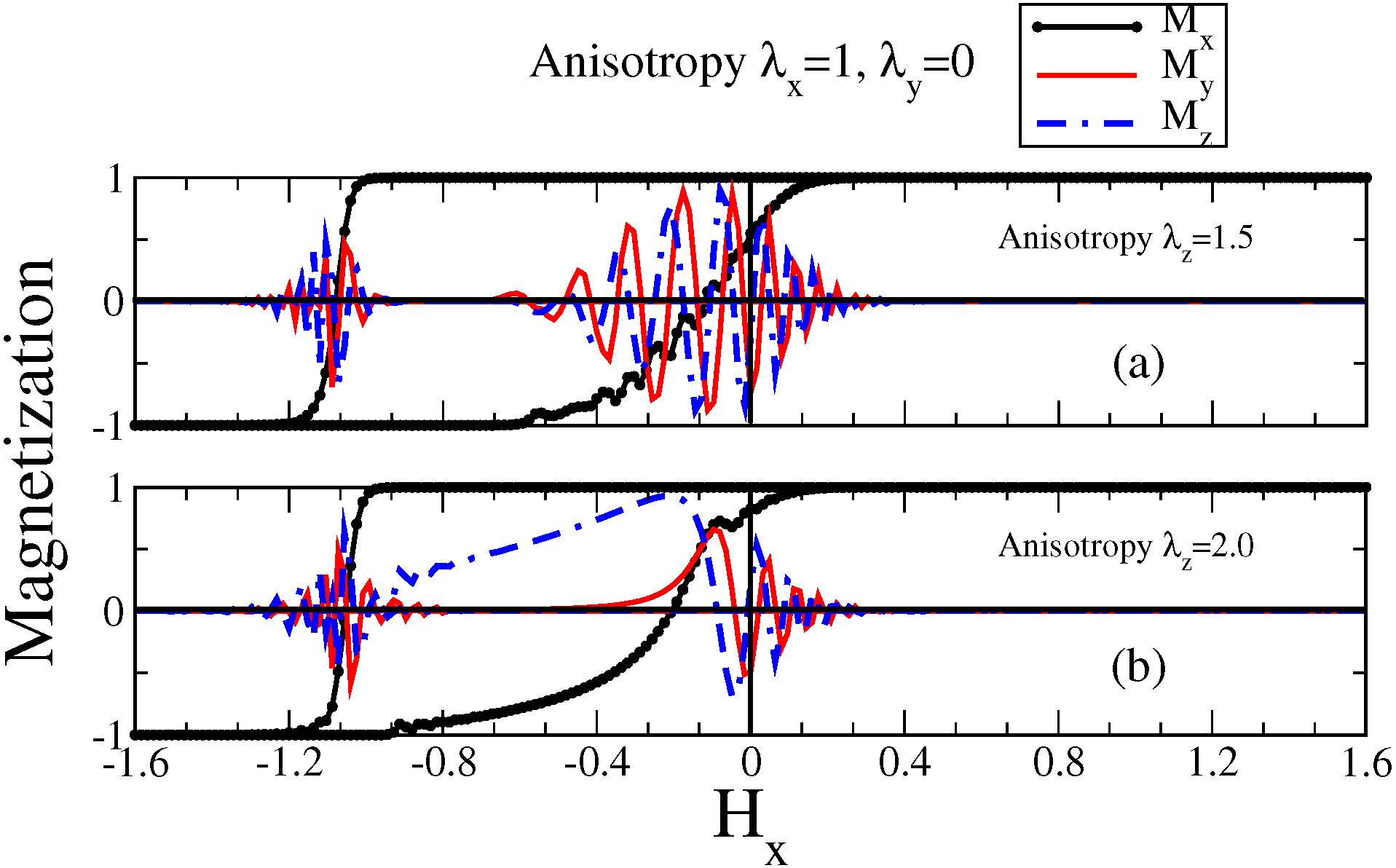}
\caption{Magnetization in the directions $x$, $y$ and $z$. The values of the anisotropy are shown in the figure. In this case, we change the anisotropy in the $z$-direction. The behavior is similar to the previous case but the precession is around the $y$-direction.}
\label{histereses_direcoes_lamby}
\end{figure}
%%%%%%%%%%%%%%%%%%%%%%%%%%%%%%%%%%
%%%%%%%%%%%%%%%%%%%%%%%%%%%%%%%%%%
\begin{figure}[h!]
\centerline{\resizebox*{8.5cm}{!}{\includegraphics{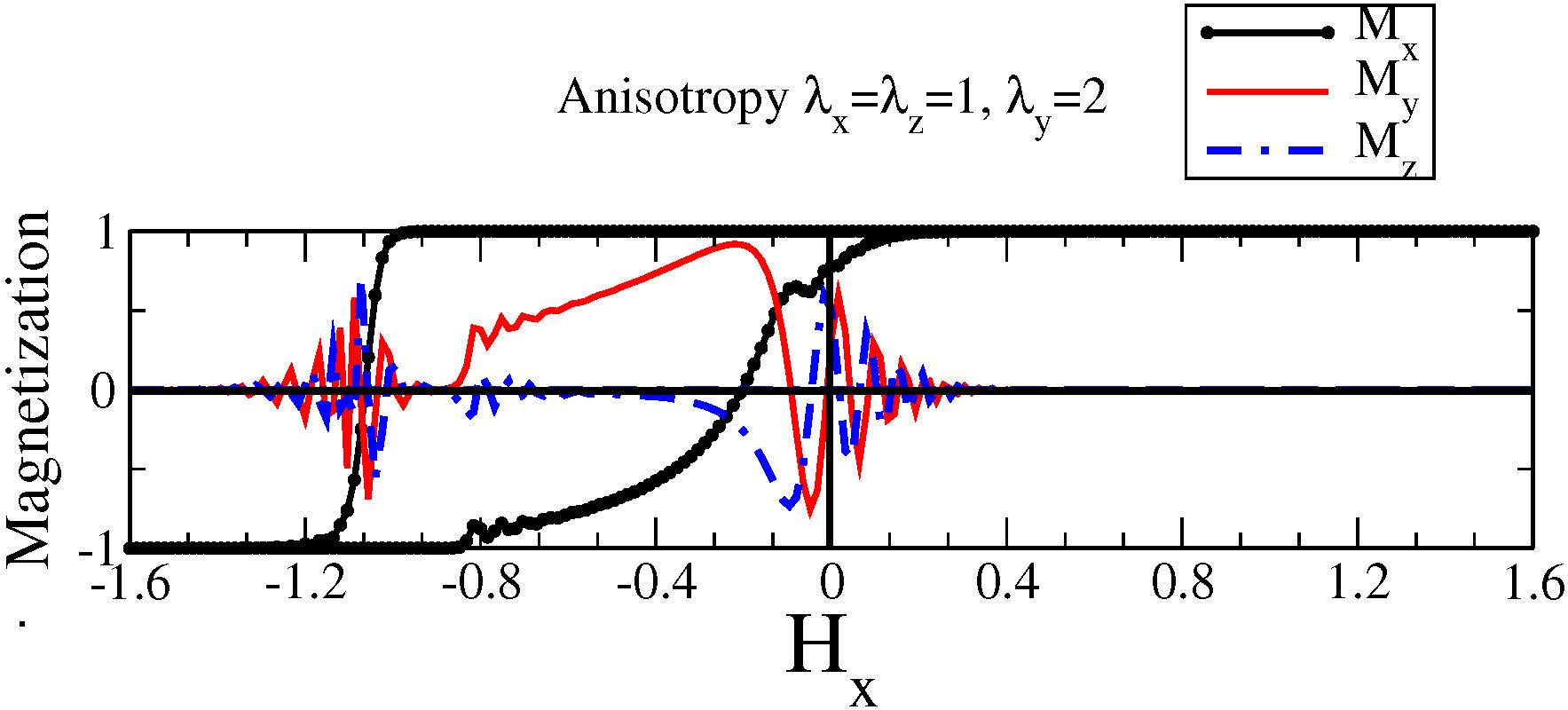}}}
\caption{Magnetization in the directions $x$, $y$ and $z$ to $\lambda_{x} = \lambda_{z} =1$ and $\lambda_{y} = 2$. The magnetization behavior is similar to the previous cases. The precession is around the $z$-direction.}
\label{histereses_direcoes_lambxy}
\end{figure}
%%%%%%%%%%%%%%%%%%%%%%%%%%%%%%%%%%
Second, to $\lambda_{y} \geq 1.63 $ (figure \ref{histereses_new_163_lambz}), the component $M_y$ almost reaches the maximum value of saturation and  and oscillates around  the $z$-direction in the ascending magnetization before the point $M_x=0$ instead of oscillates around the $x$-direction as in the case of the figure \ref{histereses_new_lambz}. In this case the dynamic behavior of the hysteresis loop can be separated in a precession around the $x$-direction (region close to the point $I$ and $III$ of the figure \ref{histereses_new_163_lambz}) and a precession  around the $z$-direction (region $II$ of the figure \ref{histereses_new_163_lambz}). In both cases, the change in coherent magnetization precession is responsible for  asymmetric hysteresis loop \cite{jmccord,jingchen,yangliu,jcamarero,egirgis,johannes,mr,beck,beck1}. Investigations has shown that the magnetic hysteresis can be either by coherent rotation or by the nucleation and growth of reverse domains (domain-wall motion). When the hysteresis is a mixture those processes the  symmetry of the hysteresis  loop is broken. In our simulation this asymmetry comes just from anisotropy in the interface that change the local field in the  AFM/FM interface causing a variation of the direction of the axis about which the magnetic moment can rotate in the ascending magnetization. In the descending magnetization, the local field in the $x$-direction is strong enough to does not alter the precession.
\\
In the figure \ref{histereses_direcoes_lamby} is shown the magnetization to an anisotropy  in the $z$-direction. In this case, the precession is around the $y$-direction in the ascending magnetization and the magnetization behavior is similar to the previous cases. To $\lambda_x=\lambda_z=1$ and $\lambda_{y}=2$ the magnetic behavior is similar to that found in the figure \ref{histereses_direcoes_lambxy}.
\\
\section{Conclusions}
We have made spins dynamics simulations considering an anisotropic exchange energy between FM/AFM layers. We verified that a strong anisotropy along a given direction  induces  a torque on the magnetics moments that rotates the magnetization field directly toward the effective magnetic field  affecting the precession rate, thereby causing the magnetization field to move directly into alignment with the effective field more and more rapidly until rotating them towards that direction. Therefore,  a magnetization rotation is induced by the addition of an anisotropy in the FM/AFM  interaction  affecting the symmetry of the hysteresis loop. 
\\
\section*{Acknowledgements}
This work was partially supported by CNPq, CAPES, FINEP and FAPEMIG (Brazilian Agencies). Numerical work was done at the Laborat\'orio de Computa\c{c}\~ao e Simula\c{c}\~ao do Departamento de F\'isica da UFJF.
%------------------------------Bibliography --------------------------------
%
%\newpage

%

\end{document}